\documentclass{mn2e}
\pdfoutput=1
\usepackage{graphicx}
\usepackage{longtable}
\usepackage{float}
\begin{document}

\title[Comment]{A dearth of dark matter in strong gravitational lenses}

\author[R.H. Sanders] {R.H.~Sanders\\Kapteyn Astronomical Institute,
P.O.~Box 800,  9700 AV Groningen, The Netherlands}

 \date{received: ; accepted: }

\maketitle
\begin{abstract}

I show that the lensing masses of the SLACS sample of strong
gravitational lenses are consistent 
with the stellar masses determined from population synthesis
models using the Salpeter IMF.  This is true in the context of 
both General Relativity and modified Newtonian dynamics, and
is in agreement with the expectation of MOND
that there should be little classical discrepancy 
within the high surface brightness regions probed by
strong gravitational lensing.  There is also dynamical
evidence from this sample supporting the claim that
the mass-to-light ratio of the stellar component 
increases with the velocity dispersion. 
\end{abstract}

\section{Introduction: mass discrepancies in elliptical galaxies}

Modified Newtonian dynamics, MOND, (Milgrom 1983) is a 
non-relativistic theory that posits
the existence of a critical acceleration ($a_0\approx 10^{-8}$
cm/s$^2$)
below which the effective gravitational acceleration $g$ 
deviates from
Newtonian form ($g_N$) in a specific way -- in effect,
$g \rightarrow \sqrt{a_0 g_N}$ when $g<a_0$.
The motivation is to remove the need for dark matter in
gravitationally bound astronomical systems with low
internal and external accelerations ($g<a_0$).  The critical acceleration
can also be expressed as a surface density ($\approx a_0/G$),
the implication being that discrepancies between the
classical dynamical mass and the observable baryonic mass
should appear
in low surface density, or low surface brightness,
systems.  Conversely, there should be no significant discrepancy
within high surface brightness systems

Rotation curves of the neutral gas in disk galaxies
as measured in the 21 cm line present obvious advantages
in tracing the gravitational acceleration as
a test of MOND:  cool gas,
generally in planar circular motion, provides
an unambiguous tracer of the acceleration and deviations from
such motions can usually (but not always) be identified; moreover,
random motions are small and generally do not contribute to
the support of the gas disk against gravity (Trachternach et al.
2008). 

The success of MOND when confronted by the extensive 
body of data on measured rotation curves, 
ranging from gas rich low surface brightness dwarfs
(Swaters et al. 2010) to high surface brightness earlier type 
disk galaxies dominated by a stellar component  
(Sanders and Noordermeer 2007), can hardly be disputed.  
This success, plus the 
observed and theoretically predicted baryonic Tully-Fisher relation
(McGaugh 2005) constitute the principal evidence supporting
MOND.   However,
gas poor early type systems, ellipticals and S0s, usually miss 
such a clear tracer of the gravitational acceleration;  for
such objects the situation has been more confused.  

An unavoidable prediction of MOND is that in
high surface brightness systems, such as luminous 
elliptical galaxies, there should be little discrepancy between the
detectable baryonic mass and the Newtonian dynamical mass within the
bright luminous object.  In other words, with the
traditional Newtonian analysis, there should be
no evidence for dark matter within the projected
radius containing half the flux of visible light, the effective
radius.  This, in fact, was the result of 
the observational study of Romanowsky et al. (2003).
They used the observed kinematics of bright planetary nebulae as
a tracer of the mass distribution in three nearby elliptical
galaxies, and  found that the results were consistent
with no substantial dark matter contribution within four
effective radii, a result shown by Milgrom and Sanders (2003) to
be in agreement with the expectations of MOND.

The use of such stellar tracers suffers from the ambiguity 
introduced by the uncertain distribution of stellar orbits,
but Milgrom (2012) has recently demonstrated that the 
pressure distribution of the hot X-ray emitting gaseous envelopes
in two isolated elliptical galaxies, extending out to 100
kpc and over a range of a factor 100 in acceleration, 
is entirely consistent with the run
of gravitational accelerations calculated from the
observed distribution of visible stars using the
MOND algorithm.  The implied mass-to-light ratios of the stellar
populations are sensible for early type galaxies.  
Subsequently, Milgrom (2013)
pointed out that the statistics of galaxy-galaxy 
weak gravitational lensing
(the small distortions in the shapes of background
galaxies in the field of nearer foreground galaxies), which
probes accelerations down to a few percent of $a_0$,
implies the asymptotic velocity
dispersion of fitted isothermal spheres is related to
the baryonic mass of the deflecting galaxies as
$M \propto \sigma^4$, exactly as required by MOND -- an
elliptical galaxy equivalent to the baryonic Tully-Fisher relation
in spiral galaxies.  In other
words it appears that the dynamics of the 
outer ``halos", more than one hundred kiloparsecs in
extent, is determined by the small fraction of baryons in the 
very center, a very strange fact indeed when viewed in the context of
dark matter.

On the other hand,
there are persistent claims that strong gravitational lensing,
the formation of multiple images or Einstein rings of
background sources by foreground galaxies, require the presence
of substantial quantity of dark matter within one or two effective radii 
in early type galaxies.  If true, this would appear to be in
contradiction to the predictions of MOND because strong lensing
can only occur in the high acceleration regime (see the discussion below).
There has been controversy about this issue in the literature, with some
authors claiming that no dark matter is required (Chiu et al. 2008,
Chen and Zhao 2008, Sanders and Land 2008, Chiu et al. 2011), while others
others argue strong lensing requires that substantial fraction of 
the total mass (up to 80\%)
is dark within two effective radii 
(Mavromatos et al. 2009, Ferreras et al. 2009, Leier et al. 2011,
Ferreras et al.  2012).

This problem is complicated by the possibility of contamination -- 
whether or not distant lenses are truly 
isolated or lying within groups or clusters --   
and by the uncertainty
of the mass-to-light ratio of the underlying stellar population.
With respect to this second problem -- that of the stellar M/L -- 
there is recent evidence,
spectroscopic and dynamical, that the initial mass function of
stars formed in early-type galaxies, the IMF,
is not universal, as is often supposed, but
becomes increasingly bottom heavy -- weighted toward
lower mass stars -- in higher mass galaxies.  That is to say,
the IMF is better described by 
that of Salpeter (1955) rather than that of Chabrier(2003) 
in systems with higher velocity dispersion or total mass.   
This in turn implies that the total stellar mass-to-light ratio 
is higher in more
massive galaxies (there is about a factor of two difference
in M/L between model
populations constructed with the Chabrier vs. Salpeter IMFs).  

Spectroscopic evidence for an increasingly dominant contribution of dwarf
stars has been given by Conroy and van Dokkum (2012),
Smith et al. (2012) and Spiniello et al. (2012, 2013).
Conroy et al. (2013) have provided dynamical evidence that in compact 
elliptical galaxies (in which dark matter presumably does not
dominate) the stellar M/L increases
systematically with galaxy velocity dispersion.  The 
evidence of Spiniello et al. (2013), using spectroscopic 
tracers of low mass stars, similarly suggests an
almost linear increase in the stellar M/L with velocity
dispersion.  

Here, in view of these developments,
I reconsider the question of whether or not
the MOND lensing masses of strong lens systems are consistent
with the stellar masses in the SLACS sample of gravitational
lenses.

\section{A comment on strong lensing with MOND}

Unlike General Relativity (GR) MOND is a non-relativistic theory.  
That means that MOND
in itself says nothing about gravitational lensing and
other relativistic effects.  There have now been several
proposed candidate relativistic extensions of MOND
(see Famaey and McGaugh 2011, for a recent review) but
there is no generally accepted theory.  In most of these 
proposals, the relationship between the deflection of
photons and the weak field force, also in the low acceleration
limit, is required to be the same as it is
in GR.  This is built into the theories and,
in fact, required by apparent coincidence of the 
classical dynamical mass of clusters of galaxies
(using galaxy kinematics or the distribution of hot gas --
non-relativistic particles)
and the lensing mass of the clusters (the effect of gravity
on photons -- relativistic particles).  Therefore, it is not
yet particularly meaningful to apply a specific relativistic
theory, such as TeVeS (Bekenstein 2004), to
the problem of lensing; the suggested interpolating
functions are just as arbitrary as those typically used
in MOND.  So here I will use one of the usual MOND
interpolating functions that works well for galaxy
rotation curves, the so-called simple form recommended
by Zhao \& Famaey (2006); i.e.,
$$\mu(x) = x/(1+x) \eqno(1) $$
with $x = g/a_0$.

The MOND algorithm is given by 
$$g\mu(g/a_0) = g_N \eqno(2)$$
where I take $a_0 = 10^{-8}$ cm/s$^2$.  Viewing MOND as
a modification of gravity, this formula
is only strictly true in spherical systems.

It can be shown that, with GR, for a spherically symmetric gravitational
lens, perfectly aligned with a small background source,
an Einstein ring will be formed if the enclosed mean surface density in the
lens exceeds 
$$\Sigma_{crit} = {{cH_0}\over {4\pi G}} F(z_l,z_s)\eqno(3)$$
where $F(z_l,z_s)$ is a dimensionless function of the
lens and source redshifts (related to the angular size distances
in units of the Hubble distance, $c/H_0$); for the lenses
considered here $F\approx 10$.

The critical surface density below which
MOND phenomenology appears is roughly
$$\Sigma_{M} \approx a_0/\pi G. \eqno(4)$$
Given that $a_0 \approx cH_0/6$ we find
that, typically 
$$\Sigma_{crit}\approx ({\pi/2})\Sigma_{M}F(z_l,z_2)\approx 15 \Sigma_M. 
\eqno(5)$$
Therefore strong lensing always occurs in the high acceleration limit.
No large discrepancy should be detected by strong gravitational
lensing (the minimum value of $F$ in a concordance cosmology
is 3.4, so it is always the case that $\Sigma_{crit}>\Sigma_M$).

That is not to say that no discrepancy whatsoever
should be present within an Einstein ring radius.  MOND can
be represented by a halo of phantom dark matter -- the
dark matter that one would presume to be present if the
MOND phenomenology were to be represented by a dark halo.
The space density of phantom dark matter is
given by
$$\rho_{pdm} = -{1\over{4\pi G}}{\nabla\cdot g} - \rho_b \eqno(6)$$
where $g$ is the MOND gravitational acceleration given
by eq.\ 2 and $\rho_b$ is the density of detectable
baryonic matter.  Asymptotically, the MOND acceleration (the
solution determined by eq.\ 2 about a point mass) goes as
$$g={{\sqrt{GMa_0}}\over r} \eqno(7)$$
then eq.\ 6 implies that in the outer regions
$$\rho_{pdm} = {1\over{4\pi}} \sqrt{{{Ma_0}\over G}} {1\over {r^2}}
\eqno(8)$$
as in an isothermal sphere.
Because this phantom halo is seen in projection, it will
contribute roughly 15\% of the total {\it projected} 
mass within an Einstein ring (the exact fraction depends upon
the interpolating function $\mu$).  

Of course, phantom dark matter is a phantom, but for determining 
the lensing properties of an object with MOND, 
the concept is useful.  For example, the MOND critical surface 
density for strong lensing is identical to that given by
eq.\ 3 {\it when the projected phantom dark matter is included};  
the true surface density of projected baryonic matter is reduced
by this same factor ($\approx$ 15\%).

\section{The SLACS sample: lensing masses vs.
stellar masses}

The SLACS lenses (Sloan Lens Advanced Camera Surveys) 
comprise a reasonably large (85), 
homogeneously selected sample of strong gravitational lenses
at redshifts typically between 0.1 and 0.3.
Most of the objects are early type galaxies, elliptical or S0,
and a number of these present almost complete Einstein rings,
simplifying the modeling and leading to quite unambiguous
lensing mass estimates inside the Einstein ring radius.
The SDSS observations also include measurements of the stellar
velocity dispersions within an aperture of three arc seconds.
Auger et al. (2009) have imaged the lensing galaxies 
with the HST in various photometric bands,    
and the effective radius of the corresponding de Vaucouleurs
profile is determined.  The broad band colors permit the 
fitting of stellar population synthesis models and, thereby,
estimates of the stellar mass.  An important free function in
these models is the form of the IMF, and Auger et al. have considered
both the popular Chabrier and Salpeter forms; they tabulate 
the estimated stellar masses of the lensing galaxies in both cases.

\begin{table*}
\center
\caption{SLACS sample lenses: observed and derived parameters}
\label{symbols}
\begin{tabular}{@{}lccccccccc}
\hline
Lens &  (1) $R_{E}$  & (2) $R_{eff}$ & (3) $M_{J}(MON)$
& (4) $M_L(GR)$ & (5) $M_{L}(MON)$ & (6) $M_*(S)(R_{E})$ 
& (7) $f_J (R_{E})$ & 
(8) $f_* (R_{E})$ & (9) $M/L_V$ \\
  & kpc & kpc & $10^{11}$  $M_\odot$ & $10^{11}$  $M_\odot$ &
 $10^{11}$  $M_\odot$ & $10^{11}$  $M_\odot$  & & & \\
\hline
0008-0004 &  6.59 & 9.45 & 6.71 & 3.60 & 2.85 & 1.86 & 0.79 & 0.52 & 5.23\\
0029-0055 &  3.48 & 7.63 & 3.02 & 1.22 & 0.99 & 1.26 & 0.81 & 1.03 & 3.13\\
0037-0942 &  4.95 & 5.66 & 5.27 & 3.00 & 2.54 & 2.58 & 0.84 & 0.86 & 3.60\\
0044+0113 &  1.72 & 4.03 & 2.78 & 0.93 & 0.86 & 0.93 & 0.93 & 1.01 & 3.74\\
0157-0056 &  4.89 & 11.10 & 6.80 & 2.68 & 2.21 & 1.78 & 0.83 & 0.67 & 3.80\\
0216-0813 &  5.53 & 11.13 &12.65 & 4.96 & 4.32 & 3.74 & 0.87 & 0.75 & 4.02\\
0252+0039 &  4.40 & 5.74 & 4.22 & 1.80 & 1.46 & 1.29 & 0.81 & 0.72 & 4.62\\
0330-0020 &  5.45 & 4.38 & 3.66 & 2.57 & 2.07 & 2.14 & 0.81 & 0.83 & 3.09\\
0728+3855 &  4.21 & 5.89 & 4.01 & 2.07 & 1.73 & 2.11 & 0.84 & 1.02 & 3.67\\
0737+3216 &  4.66 & 8.18 & 6.71 & 3.01 & 2.57 & 3.48 & 0.86 & 1.17 & 3.04\\
0819+4534 &  2.73 & 6.20 & 2.92 & 1.11 & 0.96 & 0.81 & 0.86 & 0.73 & 4.03\\
0822+2652 &  4.45 & 6.73 & 4.98 & 2.45 & 2.07 & 2.03 & 0.85 & 0.83 & 4.08\\
0903=4116 &  7.23 & 9.71 & 8.43 & 4.62 & 3.71 & 3.05 & 0.80 & 0.66 & 3.89\\
0912+0029 &  4.58 & 10.97 & 11.69 & 4.08 & 3.62 & 2.85 & 0.89 & 0.70 & 6.35\\
0935-0003 &  4.26 & 10.27 & 11.46 & 4.11 & 3.71 & 2.83 & 0.86 & 0.69 & 3.43\\
0936+0913 &  3.45 & 6.10 & 3.55 & 1.56 & 1.32 & 1.81 & 0.85 & 1.16 & 3.27\\
0946+1006 &  4.95 & 8.17 & 6.32 & 2.96 & 2.49 & 1.53 & 0.84 & 0.52 & 7.06\\
0956+5100 &  5.05 & 8.10 & 8.24 & 3.87 & 3.35 & 2.59 & 0.87 & 0.67 & 5.03\\
0959+4416 &  3.61 & 7.23 & 4.31 & 1.75 & 1.49 & 1.84 & 0.85 & 1.05 & 3.62\\
0959+0410 &  2.24 & 2.83 & 1.49 & 0.78 & 0.68 & 0.64 & 0.87 & 0.82 & 5.35\\
1016+3859 &  3.13 & 4.07 & 2.89 & 1.49 & 1.30 & 1.35 & 0.87 & 0.91 & 4.47\\
1020+1122 &  5.12 & 6.59 & 6.71 & 3.50 & 3.00 & 2.84 & 0.86 & 0.81 & 4.96\\
1023+4230 &  4.50 & 5.48 & 4.36 & 2.44 & 2.05 & 1.73 & 0.84 & 0.71 & 5.29\\
1029+0420 &  1.92 & 2.90 & 1.34 & 0.61 & 0.54 & 0.80 & 0.88 & 1.32 & 3.51\\
1100+5329 &  7.02 & 9.89 & 9.20 & 4.84 & 3.96 & 2.98 & 0.82 & 0.61 & 4.69\\
1103+5322 &  2.78 & 7.56 & 2.87 & 0.98 & 0.82 & 0.99 & 0.84 & 1.02 & 4.12\\
1106+5228 &  2.17 & 2.38 & 1.70 & 0.93 & 0.83 & 1.17 & 0.90 & 1.26 & 3.14\\
1112+0826 &  6.19 & 5.35 & 7.00 & 4.54 & 3.84 & 2.93 & 0.85 & 0.65 & 5.30\\
1134+6027 &  2.93 & 5.23 & 2.97 & 1.29 & 1.12 & 1.22 & 0.87 & 0.94 & 4.58\\
1142+1001 &  3.52 & 4.31 & 3.18 & 1.72 & 1.47 & 1.65 & 0.86 & 0.96 & 3.46\\
1143-0144 &  3.27 & 5.02 & 4.31 & 2.00 & 1.77 & 1.63 & 0.89 & 0.82 & 3.79\\
1153+4612 &  3.18 & 3.08 & 1.85 & 1.15 & 0.96 & 1.11 & 0.84 & 0.97 & 3.65\\
1204+0358 &  3.68 & 2.98 & 2.40 & 1.59 & 1.35 & 1.58 & 0.85 & 1.00 & 4.45\\
1205+4910 &  4.27 & 6.07 & 5.25 & 2.59 & 2.23 & 2.24 & 0.86 & 0.87 & 4.17\\
1213+6708 &  3.13 & 3.22 & 2.55 & 1.49 & 1.30 & 1.56 & 0.87 & 1.05 & 3.04\\
1218+0830 &  3.47 & 6.28 & 3.83 & 1.67 & 1.43 & 1.45 & 0.86 & 0.87 & 4.05\\
1250+0523 &  4.18 & 4.75 & 3.35 & 1.95 & 1.63 & 1.63 & 0.84 & 0.84 & 2.30\\
1306+0600 &  3.87 & 3.57 & 3.55 & 1.77 & 1.54 & 1.36 & 0.87 & 0.77 & 5.37\\
1313+4615 &  4.25 & 4.80 & 4.41 & 2.48 & 2.13 & 1.83 & 0.86 & 0.74 & 5.01\\
1318-0313 &  6.01 & 9.25 & 6.44 & 3.25 & 2.62 & 1.91 & 0.81 & 0.59 & 4.67\\
1330-0148 &  1.32 & 1.43 & 0.65 & 0.34 & 0.31 & 0.23 & 0.90 & 0.68 & 6.42\\
1402+6324 &  4.53 & 7.49 & 6.42 & 2.92 & 2.51 & 2.43 & 0.86 & 0.83 & 4.71\\
1403+0006 &  2.62 & 3.50 & 1.96 & 1.01 & 0.87 & 1.22 & 0.87 & 1.21 & 3.00\\
1416+5136 &  6.08 & 4.23 & 5.16 & 3.72 & 3.08 & 2.54 & 0.83 & 0.68 & 4.87\\
1420+6019 &  1.26 & 2.65 & 1.05 & 0.39 & 0.36 & 0.50 & 0.91 & 1.28 & 3.11\\
1430+4105 &  6.53 & 10.65 & 11.98 & 5.54 & 4.71 & 3.37 & 0.85 & 0.61 & 6.48\\
1436-0000 &  4.80 & 6.81 & 4.48 & 2.34 & 1.92 & 2.09 & 0.82 & 0.90 & 3.01\\
1443+0304 &  1.93 & 1.62 & 0.98 & 0.61 & 0.54 & 0.73 & 0.89 & 1.19 & 3.10\\
1451-0329 &  2.33 & 3.55 & 1.77 & 0.85 & 0.74 & 1.04 & 0.87 & 1.22 & 2.59\\
1525+3327 &  6.55 & 11.79 & 11.12 & 4.91 & 4.10 & 3.90 & 0.83 & 0.79 & 4.08\\
1531-0105 &  4.71 & 5.28 & 4.85 & 2.79 & 2.37 & 2.07 & 0.85 & 0.74 & 3.67\\
1538+5817 &  2.50 & 2.44 & 1.55 & 0.93 & 0.81 & 0.99 & 0.87 & 1.06 & 3.56\\
1614+4522 &  2.54 & 7.54 & 2.34 & 0.74 & 0.61 & 0.79 & 0.83 & 1.07 & 3.54\\
1621+3931 &  4.97 & 5.65 & 5.35 & 3.03 & 2.57 & 2.41 & 0.85 & 0.80 & 3.77\\
1627-0053 &  4.18 & 6.44 & 4.94 & 2.35 & 2.01 & 2.03 & 0.85 & 0.86 & 4.79\\
1630+4520 &  6.91 & 6.23 & 7.63 & 4.93 & 4.09 & 3.88 & 0.83 & 0.79 & 5.40\\
1636+4707 &  3.96 & 5.96 & 3.64 & 1.79 & 1.50 & 1.77 & 0.84 & 0.99 & 3.67\\
1644+2625 &  3.07 & 3.65 & 2.49 & 1.35 & 1.17 & 1.26 & 0.87 & 0.93 & 3.99\\
1719+2939 &  3.89 & 4.33 & 3.45 & 1.98 & 1.69 & 1.40 & 0.86 & 0.71 & 5.22\\

\end{tabular}
\end{table*}

\begin{table*}
\center
\label{symbols}
\begin{tabular}{@{}lccccccccc}
\hline
Lens &  (1) $R_{E}$  & (2) $R_{eff}$ & (3) $M_{J}(MON)$
& (4) $M_L(GR)$ & (5) $M_*{L}(MON)$ & (6) $M^*(S)(R_{E})$ 
& (7) $f_J (R_{E})$ & 
(8) $f_* (R_{E})$ & (9) $M/L_V$ \\
  & kpc & kpc & $10^{11}$  $M_\odot$ & $10^{11}$  $M_\odot$ &
 $10^{11}$  $M_\odot$ & $10^{11}$  $M_\odot$  & & & \\

\hline
2238-0754 &  3.08 & 4.29 & 2.61 & 1.31 & 1.13 & 1.22 & 0.86 & 0.93 & 3.80\\
2300+0022 &  4.51 & 5.39 & 5.62 & 3.04 & 2.64 & 2.09 & 0.87 & 0.69 & 5.88\\
2303+1422 &  4.35 & 7.68 & 6.23 & 2.70 & 2.32 & 1.94 & 0.86 & 0.72 & 4.87\\
2321-0939 &  2.47 & 6.17 & 3.68 & 1.23 & 1.10 & 1.21 & 0.89 & 0.98 & 4.14\\
2341+0000 &  4.50 & 7.15 & 4.84 & 2.32 & 1.94 & 2.16 & 0.93 & 0.89 & 4.00\\
2347-0005 &  6.10 & 6.11 & 7.93 & 4.76 & 4.06 & 3.46 & 0.86 & 0.73 & 3.69\\
\hline
\end{tabular}
\medskip
\begin{tabular}{@{}l}
(1) the Einstein ring radius in kpc; (2) I-band effective radius
in kpc; (3)  total Jaffe model 
mass  with modified dynamics or relativistic \\ equivalent 
such as TeVeS; (4) projected GR lensing mass within Einstein ring;
 (5) projected MOND lensing mass within Einstein ring \\ (6) 
the projected stellar mass (Salpeter) within the Einstein ring; (7)
the projected fraction 
of the MOND model mass
within the \\ Einstein ring radius with MOND (the
remainder being phantom dark matter);  (8) the projected fraction of
stellar mass within within the \\ Einstein ring,  the remainder being
Jaffe plus phantom dark matter (with MOND); (9) MOND $M/L_V$
\end{tabular}
\end{table*}

Here I have selected 65 objects from their sample.  These are
lenses which are classified as elliptical galaxies, and which
have complete photometric data, all with estimates of
the stellar mass.  In the dynamical analysis I assume that the
total light and mass distribution is given by the spherical
Jaffe model (Jaffe 1983)
with an effective radius appropriate to the particular object.
The effective radius is that provided by Auger et al. based
upon de Vaucouleurs law fits to the I-band photometry; I take this to be representative
of the true distribution of starlight and stellar mass
(the scale length of the corresponding Jaffe model is
given by $R_{J} = 1.31 R_{eff}$).

Given the numerical value of $F(z_l,z_s)$ (calculated
in the context
of the standard ``concordance" cosmology which the proper theory
of MOND should reproduce) and the effective
radius in each case,
I adjust the mass of the
Jaffe model, for both GR and MOND, in order to 
match the Einstein ring radius.  That is the 
radius (also given by Auger et al. 2009) within which the enclosed surface density, is equal the critical surface density (eq.\ 3) which, in the case
of MOND, includes projected phantom dark matter (eq.\ 6).  
The projected Jaffe model mass within the Einstein ring radius 
is the lensing mass 
for GR or MOND, but with MOND the lensing mass is, on average, 15\% lower
because of the higher effective gravitational force.
I emphasize again that the difference between the GR and MOND lensing
mass is equal to the contribution of projected phantom
dark matter.

It is the lensing mass in both cases that I will 
compare with the stellar
mass projected within the Einstein ring assuming 
that the Jaffe model with the
I-band effective radius describes the luminosity density.  
One could alternatively
choose to compare the total Jaffe mass either for GR or MOND 
with the total stellar mass, but 
this obscures the fact that strong lensing provides
only a measurement of the projected mass within the Einstein
ring.

The observed and derived parameters of the SLACS subsample
are given in Table 1.  Here for each lens I give 1)
the Einstein ring radius in kpc; 2)
the I-band effective radius in kpc; 3) the total mass of the Jaffe sphere required to produce the observed Einstein ring radius in the 
context of MOND (in all cases masses are given in units
of $10^{11}M_\odot$);  4) the projected mass within the Einstein
ring, the lensing mass, with GR (this is consistent with that given
by Auger et al. 2009, Table 4 column 3);  5) the projected Jaffe model mass within the Einstein
ring in the context of MOND, the MOND lensing mass 
(the difference between columns 
4 and 5 is the contribution of phantom dark matter); 6) the
stellar mass (Salpeter) projected within the Einstein
ring (this depends upon the effective radius); 7) the fraction
of the MOND mass to GR lensing mass;
8) the fraction of projected visible to GR lensing mass (ratio of column
6 to column 4) within the
Einstein ring;  9) the total MOND mass-to-light ratio in
the visible band.  Note that column 3 provides the normalization
of the Jaffe model in the context of MOND.  The Jaffe mass normalization
for GR is obtained by multiplying the MOND Jaffe mass (column 3) 
by the ratio of the
GR lensing mass to the MOND lensing mass (column 4 to column 5).

From the table we see that with MOND 
there is rather little
dispersion of the Jaffe mass fraction within the Einstein
radius (column 7):  $<f_J(R_{E})> = 0.855\pm 0.026$.  This is because
the Einstein radius depends upon the interior surface
density as does the radius of the onset of modified dynamics.
With MOND this would be the true baryonic mass in terms of the
estimated GR lensing mass.  
On the other hand the fraction of stellar mass to GR lensing
mass, $f_*(R_{E})$, shows a
greater dispersion, with $<f_*(R_{E})> = 0.84\pm 0.19$.  The
dispersion in this quantity,
also given by Auger et al. 2009 (Table 4, column 5), reflects
the errors, random and systematic, in the population synthesis
models for the stellar mass.  
In general the results for $f_*(R_{E})$ given here
correlate with those of Auger et al. (2009) although their
mean value is lower ($<f_*> = 0.73 \pm .19$) implying a larger
fraction of dark matter.  This is
because of the use here of the I-band
$R_{eff}$ as opposed to the V-band by Auger et al.  The
effective radii in V are generally larger which means a smaller
stellar mass projected within the fixed Einstein ring radius.  
I comment on this in the final section.  

The lensing mass with GR (column 4)
is plotted against the projected stellar mass (Salpeter) 
in Fig. 1, and the   
MOND lensing mass against the projected stellar mass in Fig. 2.  
As noted above, the MOND lensing mass is lower because of the
enhanced deflection due to the larger effective gravitational
force.  The quantities being plotted here are completely
independent: the stellar mass is determined from stellar population
models based upon the observed colors of the lenses, and the
lensing masses are determined by the Einstein ring radius with the
assumed law of gravity.

A quantitative measure of the discrepancy in these objects would be
the ratio of the projected to stellar mass.
For Newton, the mean value of this ratio over the sample
is 1.21 $\pm$ 0.27; with MOND this is 1.03 $\pm$ 0.22.
It is evident that both determinations of the dynamical mass
are consistent with each other and with the presence of no
discrepancy, as MOND would predict. 

When the stellar masses are
estimated using the Chabrier IMF, there is an
apparent discrepancy with the mean ratio of MOND lensing to stellar
mass being 1.90 $\pm$ 0.44.  This 
demonstrates the importance of the assumed IMF in
assertions about the contribution of dark matter within the inner
regions of elliptical galaxies:  a factor of two difference
between estimated lensing and stellar masses cannot be
taken as evidence for dark matter. 

\begin{figure}
\begin{center}
\includegraphics[height=7cm]{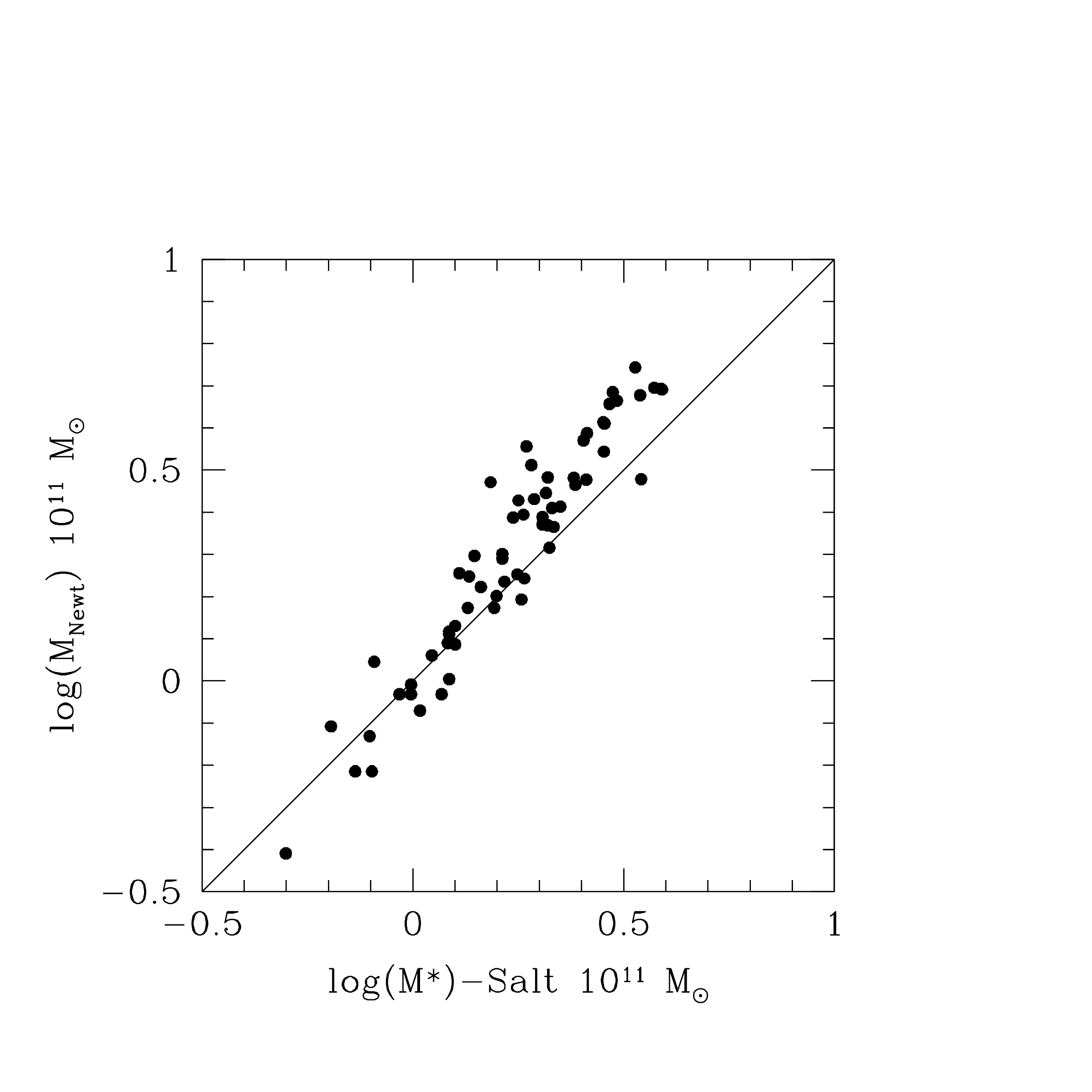}
\caption{The logarithm of the GR lensing mass
of SLACS lenses plotted against
that of the stellar
mass projected within the Einstein ring assuming the Salpeter IMF (given by Auger et al. 2009). 
Both are given in units of $10^{11}$ M$_{\odot}$  and the equality
line is shown.}
\end{center}
\end{figure}

\begin{figure}
\begin{center}
\includegraphics[height=7cm]{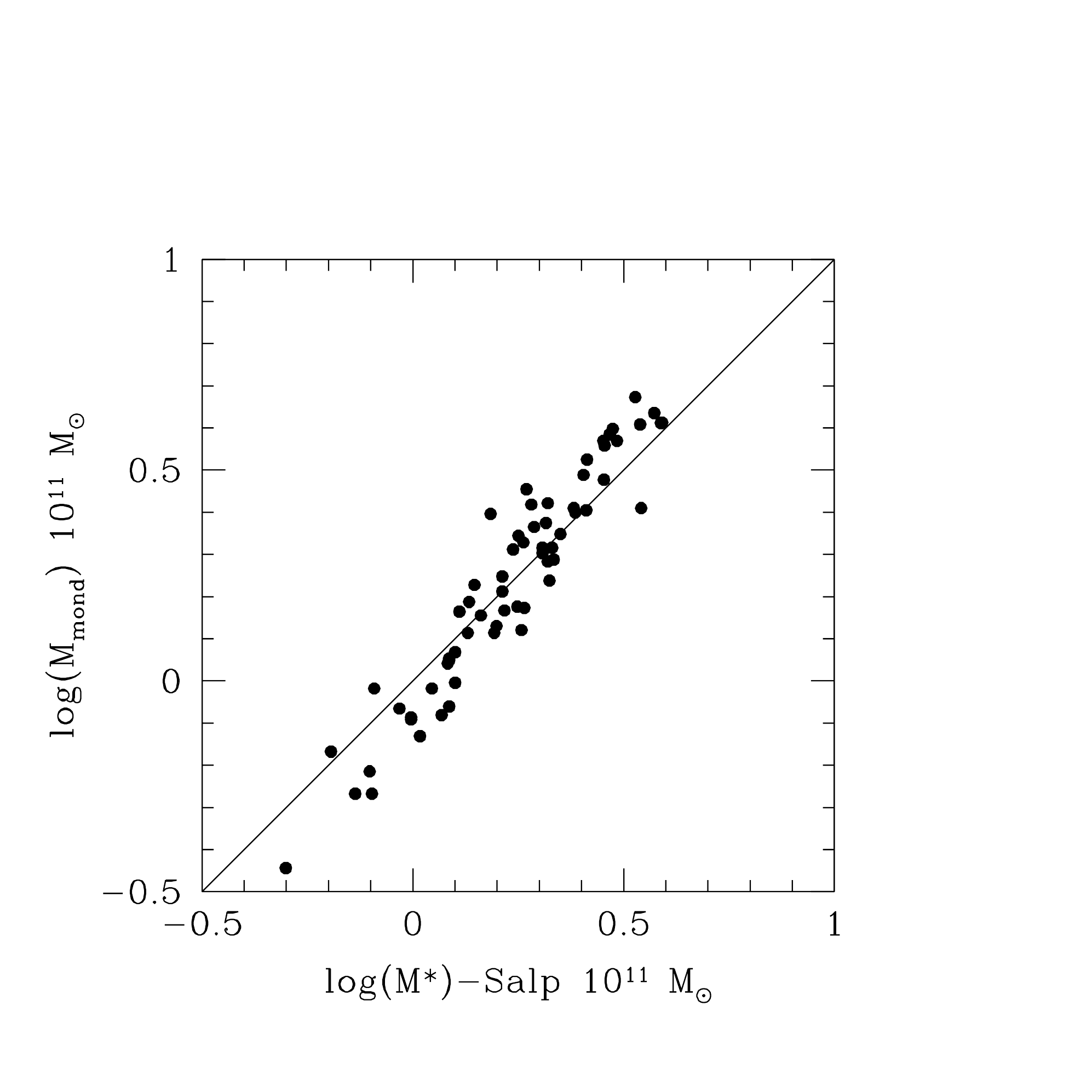}
\caption{As in Fig.\ 1 but here the logarithm of the MOND lensing mass
is plotted against the stellar mass (Salpeter) 
projected within the Einstein ring.}
\end{center}
\end{figure}

Several of the objects included here (Table 1) have been considered
in detail in separate studies.  For example SDSS J1430+4105
was discussed by Eichner, Seitz \& Bauer (2012) who argued that
several subcomponents in the lensed image  
constrain on the total mass distribution within the
Einstein radius and that the fractional dark mass within the Einstein 
ring range ranges from 0.2 to 0.4 ($0.6< f^*(R_{E})<0.8$).
The lens is complicated by the fact that it is not isolated;
there is a surrounding group.  None-the-less, the results
are roughly consistent with those given here with $f_* = 0.61$.   
The total
$M/L_V$ of this object (with modified dynamics) is the second largest of
the sample (6.48), but it is certainly not an outlier from the 
distribution of points on Fig.\ 2.

We see from Figs. 1 and 2 that there is some evidence
for a larger discrepancy at larger galaxy stellar masses
(the points appear to form a steeper relation than the 
equality line).  Indeed
there are claims that the IMF systematically varies with
galaxy mass or velocity dispersion toward being more bottom 
heavy; i.e. there is a larger stellar M/L in more massive
systems (Spiniello et al. 2013).  There is support for
this claim in Fig. 3: here the MOND lensing 
mass-to-light ratios are plotted against the observed stellar 
velocity dispersions. 
The horizontal line is the mean M/L$_V$ in the rest
frame of 4.2 $\pm$ 1.0. The points with error bars
are the M/L averaged
in bins of 15 objects and the line is the relation suggested
by Spiniello et al. on the basis of spectroscopic tracers of
low mass stars.  We see that the distribution of MOND
lensing mass-to-light ratios is reasonable for
elliptical galaxy stellar populations and that there
is marginal evidence for an increase in M/L with velocity dispersion.

\begin{figure}
\begin{center}
\includegraphics[height=7cm]{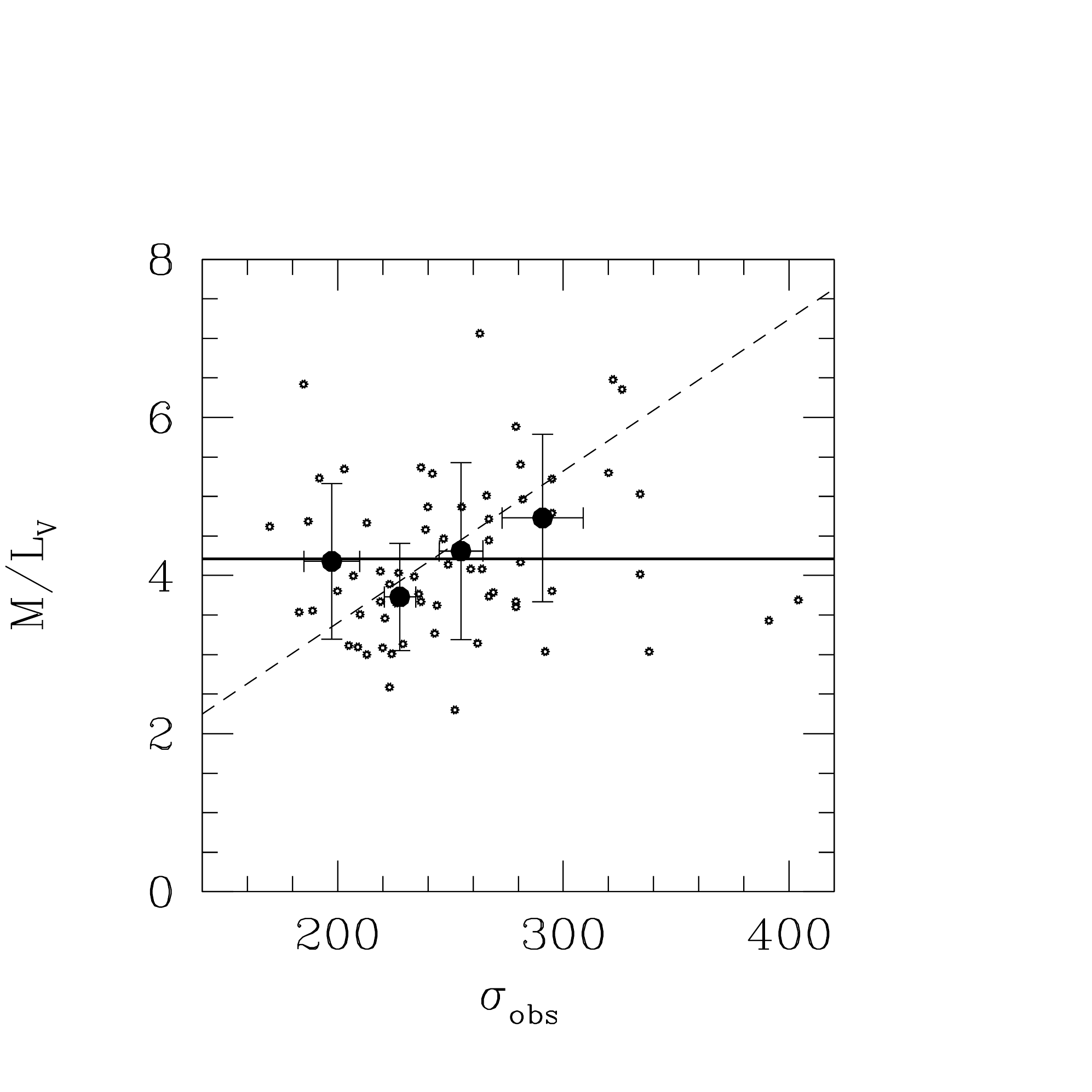}
\caption{The MOND mass-to-light (visual band) ratios of the
SLAC lenses in the rest frame plotted against the
observed central velocity dispersions.  The solid
horizontal line is the average M/L (4.2 $\pm$ 1.0), 
and the large
points with error bars are the mean M/Ls averaged
in bins of 15 objects.  The dashed line is the 
fit to stellar M/L as a function of velocity dispersion
given by Spiniello et al. 2013.}
\end{center}
\end{figure}

\section{Assessment}

Statements about the need for non-baryonic
dark matter within the bright visible inner regions
of strong gravitational lenses -- early type galaxies -- 
are not supported by
the evidence given here.  Figs. 1 and 2 demonstrate
that the GR and MOND lensing masses are
consistent with each other and with the  
masses of the stellar components determined 
from population synthesis modeling using the 
Salpeter IMF.  This
is in agreement with the expectation from MOND:
there should be little discrepancy between the visible
and the Newtonian lensing mass within high
surface brightness early type systems; the discrepancies
only appear in the outer regions.  It is also evident that the 
uncertainties introduced by the assumed IMF are 
at least a factor two.  Within this factor, no assertion about the need
for dark matter based upon use of a particular 
stellar IMF can be credible.

These conclusions are quite independent of the way in which
the systems are modeled.  Use of Hernquist (1990) rather than Jaffe
models give similar results but with slightly higher ratios of
lensing mass to stellar mass (about 6\% on average).  
The Hernquist model
does have the advantage that 
in a given object the radial distribution of stellar 
velocity dispersion 
is more nearly constant (isothermal) as observed. Since
the two classes of models bracket the empirical 
de Vaucouleurs law (Sand, et al. 2004), 
for which the effective radius is
measured, the use of an exact $r^{1/4}$ model would
certainly lie within this range of 6\%.

It is interesting that Auger et al. (2010) claim that these
observations do imply the existence of dark matter within
the inner parts of ellipticals.  There are two reasons for
the difference with the conclusions of the present work.  
The first is that
for Auger et al. the bench mark for defining the inner regions is
one-half the effective radius, whereas here I take the
Einstein ring radius.  With respect to MOND this is more appropriate 
because the onset of modified dynamics (or the appearance of
``dark matter") is tied to the enclosed
surface density as is the location of the 
Einstein ring.  Secondly, Auger et al. use the V-band
effective radius, whereas here the I-band is taken as
the indicator of the distribution of light and
stellar mass.  Which ever is more appropriate, the differences
in the estimated stellar mass within the Einstein ring are
not large and within the uncertainties of population
synthesis modeling.

Most of the claims of need for dark matter within the
Einstein radius are based upon observations of more
distant systems, such as those of the CASTLES sample
(Ferreras et al. 2012).  These lenses have a wide distribution
of redshifts, with 
typically $z_l \approx 0.4-0.5$ but ranging up to $z_l\approx 1$.
These are generally more complicated lens systems with multiple
images and relatively few complete Einstein rings; therefore
the lens modeling is less certain.  Moreover, for these distant lenses 
it is more difficult to detect contamination by
background objects -- groups or clusters surrounding the lens or
along the line of sight -- and it does appear that the
contamination rate is higher for this sample (Leier et al. 2011). 
It would seem to require substantial structural evolution
since $z=1$ if the
lenses in this sample really do have more dark matter in the
inner regions than the objects in the relatively nearby 
homogenous and clean SLACS sample. 

Overall, the MOND prediction of no significant 
discrepancy between the Newtonian dynamical mass
and stellar mass
within the inner high surface brightness regions is supported
by this analysis of the SLACS sample.  
Indeed, one can turn the argument around:
Figs. 1 and 2 support the validity of stellar 
population synthesis models in determining the mass of the
stellar component.

I thank Leon Koopmans, Chiara Spiniello, Moti Milgrom, Stacy McGaugh, 
and Marcel Pawlowsky for helpful comments.  An anonymous referee
suggested numerous improvements in the presentation of these results.

\end{document}